# Orbital dependent Coulomb drag in electron-hole bilayer graphene heterostructures

Zuocheng Zhang[1,2,*,†], Ruishi Qi[1,3,*], Jingxu Xie[1,3,4], Qize Li[1,4], Takashi Taniguchi[5], Kenji Watanabe[6], Michael F. Crommie[1,3], Feng Wang[1,3,7,†]

[1] Department of Physics, University of California, Berkeley, CA 94720, USA.

[2] Department of Physics and Astronomy, University of Nebraska-Lincoln, Lincoln, NE 68588, USA.

[3] Materials Sciences Division, Lawrence Berkeley National Laboratory, Berkeley, CA 94720, USA.

[4] Graduate Group in Applied Science and Technology, University of California, Berkeley, CA 94720, USA.

[5] Research Center for Materials Nanoarchitectonics, National Institute for Materials Science, 1-1 Namiki, Tsukuba 305-0044, Japan.

[6] Research Center for Electronic and Optical Materials, National Institute for Materials Science, 1-1 Namiki, Tsukuba 305-0044, Japan.

[7] Kavli Energy NanoSciences Institute, University of California Berkeley and Lawrence Berkeley National Laboratory, Berkeley, CA 94720, USA.

* These authors contributed equally to this work

† Correspondence to: zzhang113@unl.edu; fengwang76@berkeley.edu

**Abstract**:

We report Coulomb drag studies in an electron-hole bilayer graphene heterostructure in a magnetic field, where the orbital, spin, and valley degrees of freedom are lifted by the combined effects of exchange interaction, Zeeman energy, and vertical displacement field. Our device enables the application of a large vertical displacement field in both layers. In addition to the well-established strong Coulomb drag between Landau levels with an orbital quantum number $N = 0$, we observe a Coulomb drag signal between the $N = 1$ Landau levels under a suitable vertical displacement field. As the displacement field increases further, the Coulomb drag signal between $N = 1$ Landau levels weakens, and a Coulomb drag signal emerges between the $N = 0$ and $N = 1$ Landau levels. These findings suggest the important roles of the orbital index and vertical displacement field in interlayer Coulomb interactions within the quantum Hall regime of coupled bilayer systems.

**Main text**:

Electron-hole double-layer heterostructures, consisting of a two-dimensional electron gas (2DEG) and a two-dimensional hole gas (2DHG) separated by a nanometer-scale insulating layer, provide an ideal platform for exploring interlayer Coulomb interactions [1–3]. In the presence of a strong perpendicular magnetic field, the band structure of free electrons is quantized into Landau levels (LLs), where each state is characterized by an orbital quantum number N that governs the spatial structure of the wavefunction. This quantization quenches the kinetic energy of charge carriers, leading to a regime where interlayer Coulomb interactions dominate [4]. Coulomb drag measurements, which probes the interlayer Coulomb interaction at low temperatures, is a powerful tool for studying this interaction in these systems [5,6].

One of the most interesting phenomena in this system is the formation of interlayer excitons, where quasielectrons from a half-filled LL in one layer bind to quasiholes from a half-filled LL in the other layer [7–9]. In III-V semiconductor quantum wells, these interlayer excitons form between the lowest N = 0 LLs, leading to exciton condensation in a strong magnetic field [10]. Recent experiments in transition metal dichalcogenide double-layer heterostructures [11,12] have further highlighted the role of orbital index, where the Coulomb drag signal is strongest when both layers share the same orbital index number N and vanishes when N differs between layers. These experiments show the important role of the orbital index in the Coulomb drag phenomena.

Graphene-based systems offer another promising platform for studying Coulomb drag in electron-hole bilayers. In a double-layer system consisting of monolayer graphene, a strong Coulomb drag signal has been observed when both layers occupy the N = 0 LLs [13–16]. Bilayer graphene, with its eightfold degeneracy arising from spin, valley, and orbital degrees of freedom, provides an even richer platform for studying the role of orbital index in the Coulomb drag. Experimentally, interlayer excitons have also been observed between the $N = 0$ LLs, where the wavefunction is relatively localized [17,18]. On the other hand, the $N = 1$ LL in bilayer graphene exhibits a more spatially extended wavefunction, leading to distinct quantum phenomena. When the $N = 1$ LL is partially filled, it hosts even-denominator fractional quantum Hall effect [19,20], which is related to the non-Abelian statistics [21]. Although the incompressible state between half-filled $N = 1$ LLs has been reported in large-angle twisted double bilayer graphene [22], Coulomb drag experiments involving the $N = 1$ LL in electron-hole bilayers remain unexplored.

In this Letter, we investigate interlayer exciton formation involving the $N = 1$ LL in electron-hole bilayer graphene heterostructures. Our device is specifically designed to enable the application of relatively large vertical displacement fields across the two bilayer graphene sheets while maintaining doping in the contact regions. Under a sufficiently large vertical displacement field, we observe a Coulomb drag signal not only between the $N = 0$ LLs but also between the $N = 1$ LLs. As the displacement field is further increased, the Coulomb drag signal between the $N = 1$ LLs gradually disappears, and a Coulomb drag signal between the $N = 0$ and $N = 1$ LLs becomes more prominent. These findings provide new insights into orbital-dependent Coulomb interactions in the quantum Hall regime, highlighting the interplay between the LL orbital index and interlayer Coulomb interactions.

Figure 1a shows the cross-sectional design and electrostatic gate configuration of the electron-hole double bilayer graphene heterostructure. The device consists of two bilayer graphene (bigr) sheets, separated by a thin hexagonal boron nitride (hBN) insulating layer (~3.5 nm). The heterostructure is further encapsulated between the top and bottom hBN dielectric layers (each ~ 20 nm). The bottom gate is composed of few-layer graphite (FLG) and the top gate is made of gold (Au). The effective measurement area is defined by the overlapping region between the top and the bottom gates (between two white dashed lines in Fig. 1a). We chose a gate configuration that the top gate voltage also tunes the bottom contact, and the bottom gate voltage tunes the top contact. To achieve this, the gold electrode pads for bilayer graphene are positioned below (above) the bottom (top) bilayer graphene layer, effectively minimizing screening effects.

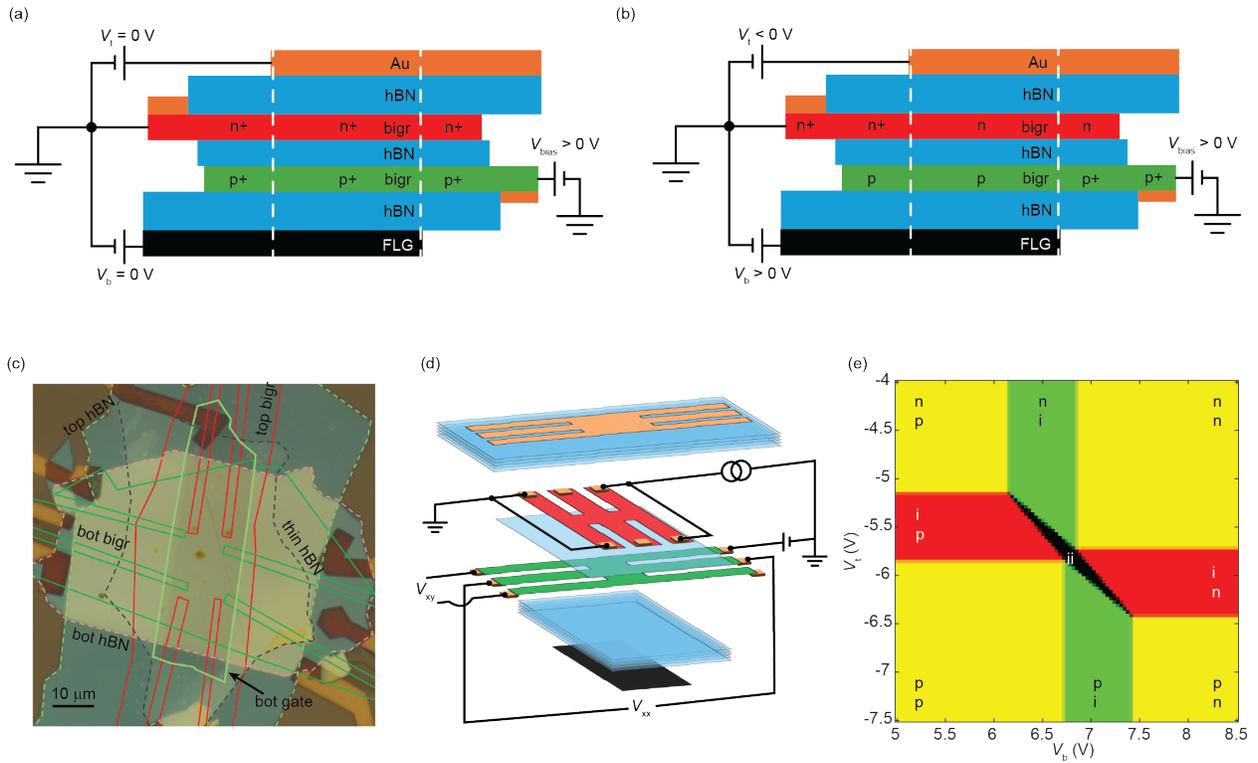

Fig. 1. (a, b) Schematic illustrations of the device structure and gate voltage configurations with the top and bottom gate voltage grounded (a) and applied (b). (c) Optical microscope image of the fabricated double-bilayer graphene device. (d) Three-dimensional schematic of the Coulomb drag measurement setup. (e) Calculated doping phase diagram of the double-layer bilayer graphene heterostructure, illustrating different carrier configurations.

Figures 1a and 1b illustrate the device operation in two steps. In Fig. 1a, both the top gate ($V_t = 0$ V) and bottom gate ($V_b = 0$ V) are initially grounded, while a positive bias voltage ($V_{bias} > 0$) is applied across the bottom bilayer graphene sheet. This configuration induces electron doping (n+) in the top bilayer graphene sheet and hole doping (p+) in the bottom bilayer graphene sheet, forming an electron-hole bilayer. Here the contact regions in both layers are nearly intrinsic as both top and bottom gates are grounded, and the top and bottom gate dielectric layers are relatively thick (bottom contact might be slightly hole-doped due to the positive bias voltage). Next, in Fig. 1b, a negative top gate voltage ($V_t < 0$) and a positive bottom gate voltage ($V_b > 0$) are applied while maintaining $V_{bias} > 0$. The application of both top and bottom gate voltages has two effects. First, it reduces the electron-hole carrier density in the central effective measurement region, tuning the system so that the top bilayer graphene is weakly electron-doped (n) and the bottom bilayer graphene is weakly hole-doped (p). Second, it affects the contact region. As the negative top gate voltage effectively injects holes into the bottom bilayer graphene contact regions, making them hole-doped (p+). Similarly, the bottom gate tunes the top contact regions, tuning top bilayer graphene contact regions to be electron-doped (n+). This contact scheme is optimized for studying electron-hole bilayers (top electron; bottom hole), but it also supports electron-electron, hole-hole, and hole-electron configurations, as the contacts remain doped across all gating conditions with a fixed bias gate voltage.

The top gate ($V_t$), bias gate ($V_{bias}$), and bottom gate ($V_b$) control the vertical displacement fields in the bilayer graphene layers through the vertical electric fields across the top hBN dielectric layer ($E_t$), middle hBN dielectric layer ($E_m$), and bottom hBN dielectric layer ($E_b$). The carrier density and vertical displacement field in the bottom bilayer graphene are [23]:

$$n_b = \frac{\varepsilon_0 \varepsilon_{hBN}}{e}(E_b - E_m), \quad D_t = \frac{\varepsilon_{hBN}}{2}(E_b + E_m)$$

The carrier density and displacement field in the top bilayer graphene are

$$n_t = \frac{\varepsilon_0 \varepsilon_{hBN}}{e}(E_m - E_t), \quad D_b = \frac{\varepsilon_{hBN}}{2}(E_m + E_t)$$

When both bilayer graphene sheets are intrinsic ($n_t = n_b = 0$) and the chemical potentials are aligned, the vertical displacement field across both bilayer graphene layers is $D = \varepsilon_{hBN} E_m = \varepsilon_{hBN} \frac{V_{bias}}{d_m}$. Here $\varepsilon_{hBN} \approx 3.3$ and $d_m$ is the thickness of the thin hBN layer $d_m \approx 3.5$ nm.

Figure 1c shows a microscope image of the device (the top gold electrode is not shown for clarity). The top bilayer graphene is outlined by red solid lines, and the bottom bilayer graphene is outlined by green solid lines. The heterostructure is assembled using a polyethylene terephthalate glycol (PETG) based dry transfer technique [24]. Each bilayer graphene sheet is precut using an atomic force microscope (AFM) tip to define six electrodes, enabling independent four-terminal measurements. The precut line lies within the effective measurement region. This design allows reliable probing of the longitudinal resistance ($R_{xx}$) and Hall resistance ($R_{xy}$) within the effective measurement area. The outlines of the top, thin, and bottom hBN layers are marked by dashed lines. The cross-stacking arrangement of the two bilayer graphene sheets facilitates fabrication of separate electrode contacts to both layers. Figure 1d illustrates a three-dimensional schematic of the device, corresponding to the microscope image in Fig. 1c. A 1 nA AC current is applied through two corners of the top layer, with the drain connected at the opposite two corners. The longitudinal resistance ($R_{xx}$) and Hall resistance ($R_{xy}$) are measured in the bottom layer using a standard lock-in technique at a frequency around 17 Hz.

Figure 1e presents the calculated doping phase diagram for the double-bilayer graphene system as a function of $V_t$ and $V_b$ under a bias voltage of $V_{bias} = 1$ V, following the theory developed in Ref. [25]. The dielectric layer thicknesses are chosen to match those of our actual device. We assume the bandgap of the bilayer graphene to be around 100 meV at $V_b = 7$ V and $V_t = -6$ V in both layers [20,23]. The red and green regions represent charge-neutral states in the top and bottom layers, respectively. In the phase diagram, each layer can be tuned from hole-doped (p) to electron-doped (n), crossing through its intrinsic charge-neutral (i) region. This results in nine distinct regions in the phase diagram, which are indicated in Fig. 1e. Among these, we focus on four doped corner regions: the electron-hole region (top left), hole-electron region (bottom right), hole-hole region (bottom left), and electron-electron region (top right). These regions are of interest for studying Coulomb drag, as the formation of LL enhances interlayer Coulomb interactions. In the center of region ii, the vertical displacement field is estimated by $D \sim \varepsilon_{hBN} \frac{V_{bias}}{d_m} = 0.94$ V/nm. Moving toward the hole-electron region (bottom right), the vertical displacement field slightly increases due to induced doping in both layers.

Figure 2a illustrates the evolution of LLs in bilayer graphene as a function of vertical displacement field at a fixed magnetic field. In bilayer graphene, the LLs exhibit eightfold degeneracy arising from orbital, spin, and valley degrees of freedom [26,27]. The vertical axis represents energy, and the horizontal axis corresponds to the vertical displacement field D. The N = 0 LL is depicted as an open ellipse, and the N = 1 LL is shown as a filled ellipse. The orange color indicates K+ valley polarization and blue color represents K- valley polarization. The up and down arrows correspond to spin polarization respectively. When the vertical displacement field is close to zero, the overlapping colors indicate no valley polarization. The energy separation between the N = 0 and N = 1 LLs is determined by exchange interactions and the energy separation between spin-up and spin-down LLs is set by the Zeeman energy. Since the Zeeman energy in a strong magnetic field (B = 10 T) is larger than the exchange energy in bilayer graphene [28], the upper four LLs correspond to spin-up states, and the lower four LLs correspond to spin-down states. As the vertical displacement field increases, the eight LLs evolve along the dashed lines. After three intersection points, the four hole LLs and four electron LLs become valley-polarized, as indicated by the distinct orange and blue colors on the right. The small schematics on the right further illustrate the wavefunction distribution corresponding to each state, which is denoted by the orbital, spin, and valley quantum numbers ($|N\sigma K\rangle$) associated with the LLs in each layer. This behavior has been established in previous studies [28–34]. We will focus on the Coulomb drag in this fully spin-, valley-, and orbital-polarized regime to avoid the complications arising from potential LL overlaps.

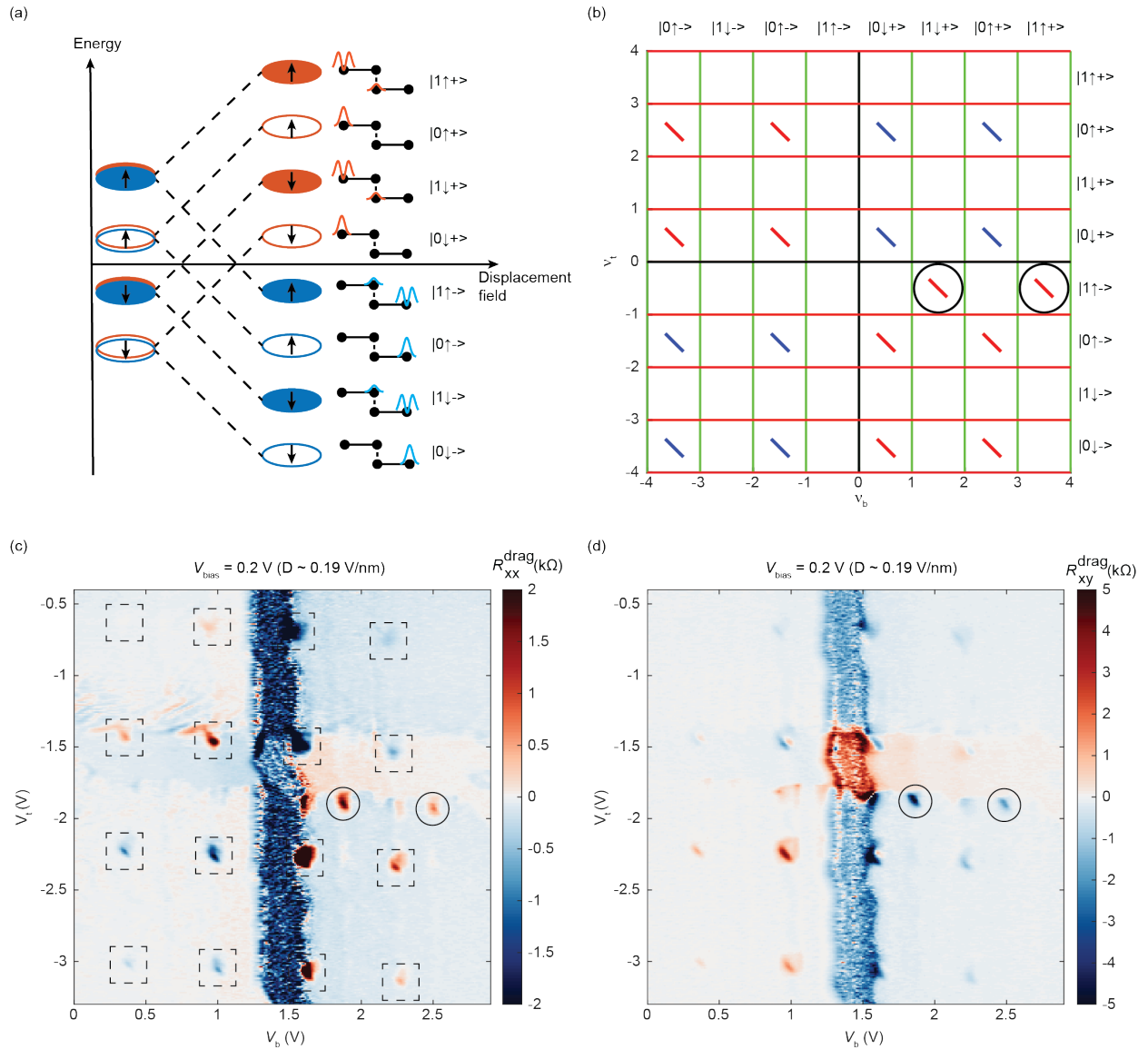

Fig. 2. (a) Schematic illustration of the Landau level structure in bilayer graphene under a displacement field. (b) Schematic representation of the expected diagonal Coulomb drag signal for different filling factor combinations ($\nu_b$, $\nu_t$) in the double-layer bilayer graphene system. (c, d) Two-dimensional color plots of the longitudinal Coulomb drag resistance ($R_{xx}^{drag}$) (c) and Hall drag resistance ($R_{xy}^{drag}$) (d) as a function of bottom gate voltage ($V_b$) and top gate voltage ($V_t$) at $V_{bias} = 0.2$ V.

Figure 2b presents the LL filling factor diagram for the double-layer bilayer graphene system, where the horizontal axis ($\nu_b$) represents the filling factor of LLs in the bottom bilayer graphene layer, and the vertical axis ($\nu_t$) represents the filling factor of LLs in the top bilayer graphene. The black gridlines at $\nu_b = 0$ and $\nu_t = 0$ indicate charge-neutral point. The red and green gridlines separate different quantized LLs in the top and bottom bilayer graphene respectively. Both axes range from -4 to +4, reflecting the eightfold degeneracy of bilayer graphene LLs. Negative values indicate hole doping and positive values indicate electron doping. Labels at the top and right edges of the diagram denote the orbital, spin, and valley quantum numbers ($|N\sigma K\rangle$) associated with the LLs in each layer, following the sequence depicted in Fig. 2a. Previously Coulomb drag signals ($R_{xx}^{drag}$) were observed between the N = 0 LLs of the top and bottom bilayer graphene [17,18], denoted by the red and blue diagonal lines within the grid. The color scale indicates the sign of drag resistance, with red representing positive drag ($R_{xx}^{drag} > 0$) and blue representing negative drag ($R_{xx}^{drag} < 0$). Below we show that a Coulomb drag signal between the N = 1 LLs of the top and bottom bilayer graphene, denoted by the circled diagonal lines in Fig. 2b, can emerge under a suitable vertical electrical field.

Figure 2c shows the two-dimensional (2D) color plot of the measured longitudinal drag resistance ($R_{xx}^{drag}$) in the double-layer bilayer graphene system as a function of the bottom gate voltage ($V_b$) and top gate voltage ($V_t$), with a bias voltage of $V_{bias}$ = 0.2 V. The vertical displacement field is estimated to be $D \sim 0.19$ V/nm when both layers are intrinsic. The magnetic field is 10 T, and the nominal temperature is around 10 mK. The color scale indicates the magnitude and sign of the drag resistance, with red representing positive drag ($R_{xx}^{drag} > 0$), blue representing negative drag ($R_{xx}^{drag} < 0$), and white indicating near-zero drag resistance. The dashed squares mark regions corresponding to the diagonal Coulomb drag signals marked in Fig. 2b. As expected, we observe 15 Coulomb drag signals between the N = 0 LLs with the correct sign. The signal at $(\nu_b, \nu_t)$ = (-3.5, 2.5) is absent. Additionally, the two circled regions in Fig. 2c highlight newly observed Coulomb drag signals. These regions exhibit pronounced positive drag resistance ($R_{xx}^{drag} > 0$) between the N = 1 LLs, specifically occurring when the top layer is hole-doped and the bottom layer is electron-doped at $(\nu_b, \nu_t)$ = (1.5, -0.5) and (3.5, -0.5). Figure 2d shows the corresponding 2D color plot of the measured Hall drag resistance ($R_{xy}^{drag}$) under the same experimental conditions. The two circled regions in Fig. 2d again highlight the newly observed negative Hall drag signals between the N = 1 LLs.

We next zoom in these two regions. Figure 3a focuses on the first region and presents a 2D color plot of the Hall drag resistance ($R_{xy}^{drag}$) as a function of the bottom gate voltage ($V_b$) and top gate voltage ($V_t$) at a bias voltage of $V_{bias}$ = 0.2 V. The dispersive blue signal in the plot corresponds to strong negative Hall drag resistance. Along the diagonal drag signal, the total LL filling factor is $\nu = \nu_t + \nu_b = 1$, where $\nu_t = -0.5$ and $\nu_b = 1.5$ at the center. The Hall drag signal gradually diminishes as the doping deviates from the center. Figure 3b shows a line cut of the Hall drag resistance along the black dashed line in Fig. 3a. Here, the Hall drag signal reaches a magnitude as large as 6 kΩ.

In the exciton condensation regime, the Hall drag resistance is theoretically expected to be quantized at $1/\nu * h/e^2$, where $h/e^2 \approx 25.8$ k$\Omega$. While the observed resistance does not yet reach this quantized value for an exciton condensate at 10 T, the robust Hall drag resistance suggests the presence of interlayer excitons in the electron-hole bilayer. The inset schematically illustrates the relevant interlayer LL interactions, indicating the negative Hall drag signals originate from interactions between the $|1\uparrow->$ state in the top layer and $|1\downarrow+>$ state in the bottom layer. The displacement field is directed upward. Similarly, Figure 3c focuses on the second region and presents a 2D color plot of the Hall drag resistance ($R_{xy}^{drag}$) as a function of the bottom gate voltage ($V_b$) and top gate voltage ($V_t$) at a bias voltage of $V_{bias} = 0.2$ V. The overall features are similar to those in Fig. 3a. Here along the diagonal blue line, the total filling factor is around $\nu = \nu_t + \nu_b = 3$, where $\nu_t = -0.5$ and $\nu_b = 3.5$ at the center. Figure 3d shows a line cut of the Hall drag resistance along the black dashed line in the 2D plot. Here, the Hall drag signal reaches a magnitude as large as 1.6 k$\Omega$. The inset schematically shows the relevant LL states involved in the Coulomb drag process. The top layer is still in the $|1\uparrow+>$ state, but the bottom layer is in the $|1\uparrow->$ state, which share the same spin as the top layer. These two dispersive Hall drag signals, one with opposite spin orientation and the other with same spin orientation, demonstrate strong Coulomb drag between the N = 1 LLs in the double-layer bilayer graphene heterostructure.

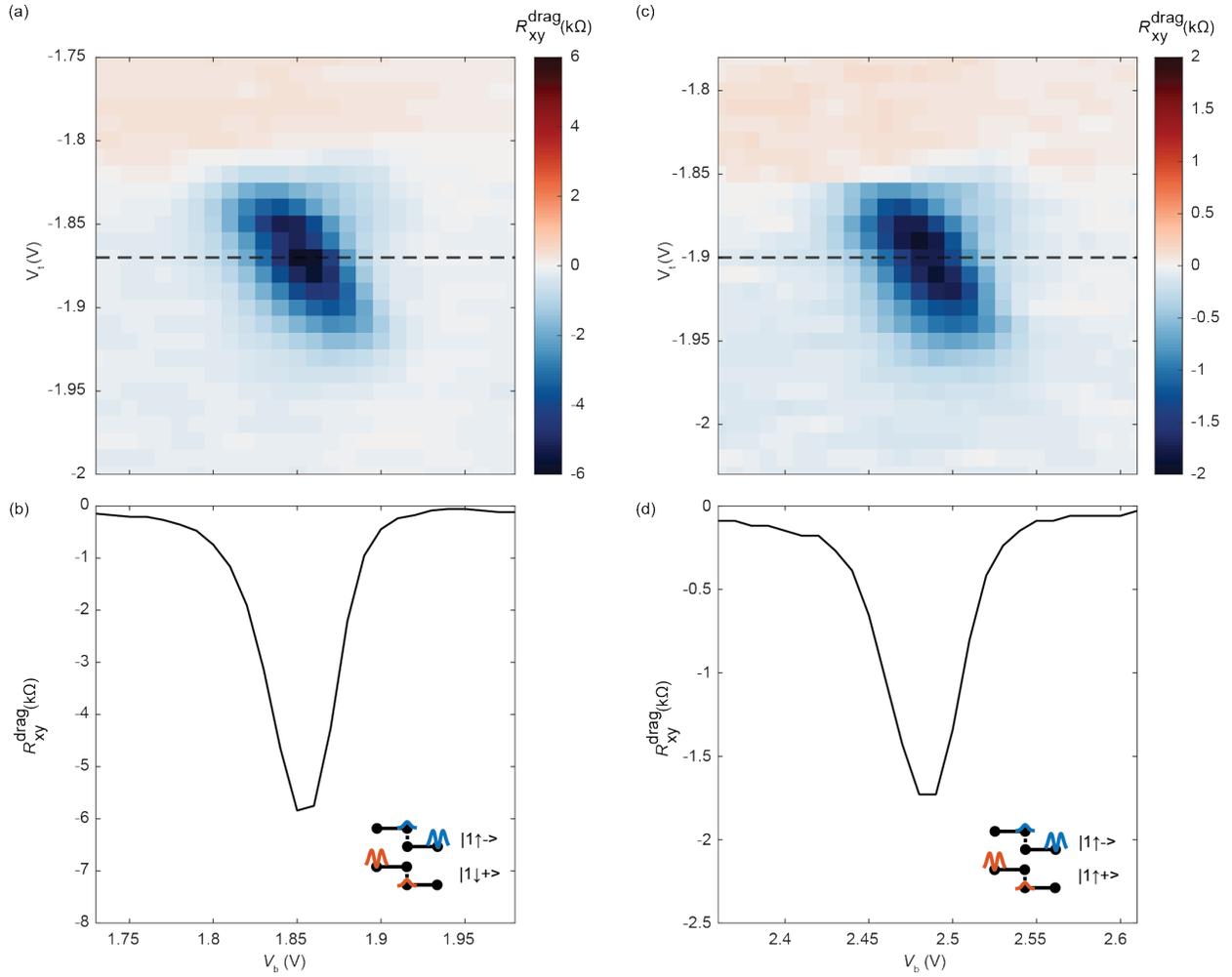

Fig. 3. (a, c) Two-dimensional color plots of the Hall drag resistance ($R_{xy}^{drag}$) as a function of bottom gate voltage ($V_b$) and top gate voltage ($V_t$) at a bias voltage of $V_{bias} = 0.2$ V. The center of each plot corresponds to the filling factor combinations ($\nu_b, \nu_t$) = (1.5, -0.5) (a) and (3.5, -0.5) (c). (b, d) Line cuts of $R_{xy}^{drag}$ taken along the dashed lines in (a) and (c). The insets indicate the relevant N = 1 Landau levels in each layer that are responsible for the observed Coulomb drag signals.

Figure 4 presents a 2D color map of the Hall drag resistance as a function of the top gate voltage ($V_t$) and bottom gate voltage ($V_b$) for various bias voltages. Each panel corresponds to a different

$V_{bias}$, ranging from 0.2 V (Fig. 4a) to 1.0 V (Fig. 4e), with the associated vertical displacement field (D) indicated in each subplot, increasing from 0.19 V/nm (Fig. 4a) to 0.96 V/nm (Fig. 4e). The selected ranges of $V_t$ and $V_b$ in each subplot ensure the filling factors remain within $-1 \leq \nu_t \leq 0$, and $0 \leq \nu_b \leq 4$. In Fig. 4a, four distinct regions of negative signal appear, corresponding to the filling factor combinations at the center $(\nu_b, \nu_t) = (0.5, -0.5), (1.5, -0.5), (2.5, -0.5)$ and $(3.5, -0.5)$. Two drag signals, at $(1.5, -0.5)$ and $(3.5, -0.5)$, were discussed in Fig. 3. Here, we first focus on the effect of the vertical displacement field on these two drag signals. As the $V_{bias}$ increases, the effective vertical displacement field increases at these points, allowing us to study the evolution of Coulomb drag with vertical displacement field. The strong Hall drag resistance observed at $(1.5, -0.5)$ and $(3.5, -0.5)$ weaken with increasing vertical displacement field. The $(1.5, -0.5)$ Coulomb Hall drag signal vanishes at $V_{bias} = 1$ V, and the $(-0.5, 3.5)$ Coulomb Hall signal disappears at $V_{bias} = 0.8$ V. Additionally, a new Coulomb drag signal at $(-0.5, 2.5)$ gradually becomes more pronounced, reaching its maximum at $V_{bias} = 0.8$ V (Fig. 4d). It represents the Coulomb drag between the N = 1 ($|1\uparrow->$) and N = 0 ($|0\uparrow+>$) LLs. Along the diagonal line, the total filling factor is $\nu = \nu_t + \nu_b = 2$. The maximum Hall drag resistance is around 2 kΩ. The dispersive quantum Hall drag resistance at $(-0.5, 2.5)$ gradually weakens as the bias voltage deviates from this critical value.

At high vertical displacement fields, LL crossing can occur on the hole-doping side between the ($|0\uparrow->$) and ($|1\downarrow->$) LLs [35]. In our drag measurements, we focus on the first LL ($|0\uparrow+>$) on the hole-doping side for the top bilayer graphene, where no such crossings are expected at high vertical displacement field. On the electron-doping side, a previous study has shown a merging tendency between the first and second LLs, as well as between the third and fourth LLs [35]. In our drag measurements, we observed four signals at the filling factor combinations $(\nu_b, \nu_t) = (0.5, -0.5)$, $(1.5, -0.5), (2.5, -0.5)$ and $(3.5, -0.5)$. We assign the third signal at $(\nu_b, \nu_t) = (2.5, -0.5)$ to Coulomb

drag signal between the N = 0 LL (The third LL on the electron-doping side in bottom bilayer graphene) and N = 1 LL (The first LL on the hole-doping side in the top bilayer graphene). However, the small energy gap between the two LL states might lead to some hybridization of the electron orbital. This change in the orbital wavefunction of LL could be related to the modified electron-hole drag behavior observed in our study.

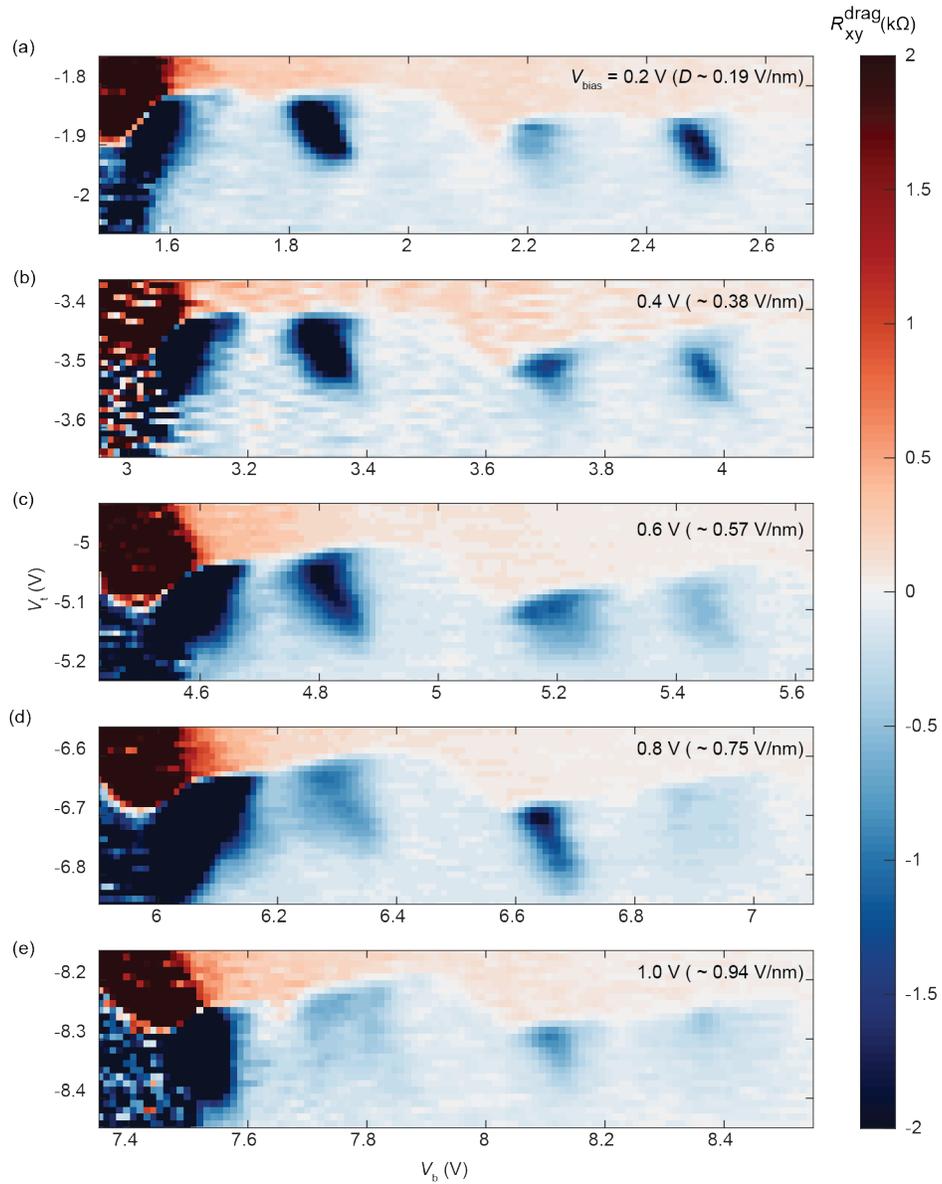

Fig. 4. (a–e) Two-dimensional color plots of the Hall drag resistance ($R_{xy}^{drag}$) as a function of bottom gate voltage ($V_b$) and top gate voltage ($V_t$) with increasing bias voltages. The applied bias voltages are $V_{bias}$ = 0.2 V (a), 0.4 V (b), 0.6 V (c), 0.8 V (d) and 1.0 V (e). A new Coulomb Hall drag signal between N = 1 and N = 0 Landau levels emerges at higher displacement fields.

In previous studies of GaAs and graphene double-layer systems, strong Coulomb drag was only observed in the N = 0 LLs. It was hypothesized that only the N = 0 orbital states can support a bilayer exciton state featuring strong Coulomb drag behavior [17]. Theoretically, bilayer quantum Hall exciton condensates correspond to easy-plane quantum Hall ferromagnetic states, and the ferromagnetic anisotropy depends sensitively on the competition between electrostatic and exchange interactions [36,37]. Our data show that strong Coulomb drag signals can exist between two N = 1 LLs and between N = 0 and N = 1 LLs, and their strengths vary significantly with the bilayer graphene bandgap tuned by a vertical displacement field. It suggests that the field-induced modification in the orbital wavefunctions might change the quantum Hall ferromagnet anisotropy and generate new bilayer exciton states between different quantum Hall orbitals. A quantitative understanding of the orbital-dependent Coulomb drag behavior at different displacement fields will require further theoretical investigations, which can shed light on the intricate competition of many-body states in electron-hole quantum Hall bilayer systems.

**Competing interests**

The authors declare that they have no competing interests.

**Acknowledgements**

The work is primarily supported by the Director, Office of Science, Office of Basic Energy Sciences, Materials Sciences and Engineering Division of the US Department of Energy under contract number DE-AC02-05CH11231 (vdW heterostructure Program KCWF16). The dilution fridge measurements are supported by the U.S. Department of Energy, Office of Science, National Quantum Information Science Research Centers, Quantum Systems Accelerator. K.W. and T.T. acknowledge support from the JSPS KAKENHI (Grant Numbers 21H05233 and 23H02052), the CREST (JPMJCR24A5), JST and World Premier International Research Center Initiative (WPI), MEXT, Japan. Z.Z. acknowledges support from NSF EPSCoR RII Track-1: Emergent Quantum Materials and Technologies (EQUATE), under Grant No. OIA-2044049. Z.Z also acknowledges support from CAS Spark Grant, Fall 2024, and Research Council Faculty Seed Grant, and startup fund from the University of Nebraska-Lincoln.

**Author contributions**

FW and ZZ conceived the research. RQ and ZZ fabricated the device and performed most of the experimental measurements together. JX and QL help with the sample fabrication and measurements. MC contributed to the fabrication of van der Waals heterostructures. ZZ, RQ and FW performed data analysis. KW and TT grew hBN crystals. All authors discussed the results and wrote the manuscript.